Title: Mathematical modeling applied to the left ventricle of heart

Authors: Saeed Ranjbar [1,*], Mersedeh Karvandi [2]

**Research institute:**

1- Modarres Hospital, Institute of Cardiovascular Research, Shahid Beheshti University of Medical Sciences, Tehran, Iran

2- Taleghani Hospital, Shahid Beheshti University of Medical Sciences, Tehran, Iran.

**\*Corresponding Author:** Saeed Ranjbar

The full postal address of the corresponding author:

Modarres Hospital, Institute of Cardiovascular Research, Shahid Beheshti University of Medical Sciences, Tehran, Iran

E-mail: Sranjbar@ipm.ir

Tel.: 00982123031331            Fax: 00982122432576

Running title: Left ventricular mathematical modeling

Abstract:

**Background:** How can mathematics help us to understand the mechanism of the cardiac motion? The best known approach is to take a mathematical model of the fibered structure, insert it into a more-or-less complex model of cardiac architecture, and then study the resulting fibers of activation that propagate through the myocardium. In our paper, we have attempted to create a novel software capable of demonstrate left ventricular (LV) model in normal hearts.

**Method:** Echocardiography was performed on 70 healthy volunteers. Data evaluated included: velocity (radial, longitudinal, rotational and vector point), displacement (longitudinal and rotational), strain rate (longitudinal and circumferential) and strain (radial, longitudinal and circumferential) of all 16 LV myocardial segments. Using these data, force vectors of myocardial samples were estimated by MATLAB software, interfaced in the echocardiograph system. Dynamic orientation contraction (through the cardiac cycle) of every individual myocardial fiber could be created by adding together the sequential steps of the multiple fragmented sectors of that fiber. This way we attempted to mechanically illustrate the global LV model.

**Result:** Our study shows that in normal cases myocardial fibers initiate from the posterior basal region of the heart, continues through the LV free wall, reaches the septum, loops around the apex, ascends, and ends at the superior-anterior edge of LV.

**Conclusion:** We were able to define the whole LV myocardial model mathematically, by MATLAB software in normal subjects. This will enable physicians to diagnose and follow up many cardiac diseases when this software is interfaced within echocardiographic machines.

## Background:

In a great many problems the microscopic structure of matter can be disregarded and a biological body replaced by a continuous mathematical model whose geometrical points are identified with material points of the body. The learning of such models is in the province of the mechanics of continuous media, which covers a vast range of problems in elasticity, hemodynamics, aerodynamics, plasticity, and electrodynamics. When the relative position of points in a continuous biological body is altered, we say that the body is strained. The change in the relative position of a point is a deformation, and the study of deformations is the region of the analysis of strain. In this paper, the left ventricular myocardium is taken as a biological elastic body and modeled by mathematical/imaging techniques.

Imaging cardiology is largely based on the evaluation of the tissue mechanics by mean of medical imaging techniques like echocardiography that has the advantage of a high temporal resolution and permits to differentiate the cardiac phases and produce an enormous amount of data about the cardiac structure and function. These data are mostly used for visualization purposes, although physically-based models would be required in order to improve their interpretation and eventually support diagnostic and therapeutic practice. A novel mathematical Cardiac modeling represents a challenge in this sense. The visualization of such this modeling in the heart chambers particularly in the left ventricle is, in fact, still of little usage in diagnosis.

Left ventricular twist deformation is due to the complex helical myocardial fiber architecture. This wringing motion of the heart is the result of the clockwise rotation of the base and the counterclockwise rotation of the apex and plays an important role in ventricular performance. Several invasive and non-invasive techniques have been used to describe and quantify this cardiac motion. Recently developed, two-dimensional speckle-tracking echocardiography (2DSTE) has proved to be a simple non-invasive technique to quantify the LV twist mechanism.[1–7] A novel technique based on three-dimensional speckle-tracking echocardiography (3DSTE) has been developed to reduce some of the shortcomings of 2DSTE.[8]

We present a new mathematical tool by introducing the notions of 2D and 3D strains which are two by two and three by three matrices. These matrices are computable from echocardiogram data. Having used these data and some new mathematical techniques and formulas we obtained can be achieved to test the hypothesis that myofibres move on helical bands which begin from the Septal, loop around the Apex and then go to the Anterior. The results we have obtained so far confirm this hypothesis. [33]

**Method and Material**

Force imaging with stimulated Matlab 7.0.4 (MathWorks, Natick, MA) software provides high spatial resolution measurement of three-dimensional myocardial points tracking (Lagrangian displacement) over the entire cardiac cycle. Echocardiography was performed on 70 healthy volunteers. Data evaluated included: velocity (radial, longitudinal, rotational and vector point), displacement (longitudinal and rotational), strain rate (longitudinal and circumferential) and strain (radial, longitudinal and circumferential) of all 16 LV myocardial segments that were prospectively acquired on Vivid E9 with 4V probe (GH Healthcare, Horton, N). [8–11] The medical ethics committee of Shahid Beheshti Medical University approved this study. All procedures were in accordance with the Helsinki declaration, and no harm was experienced by the participants. All data sets comprise a force vector field from end diastole to the end systole, when the myocardium contraction is at its maximum (Figure 1 A and B).

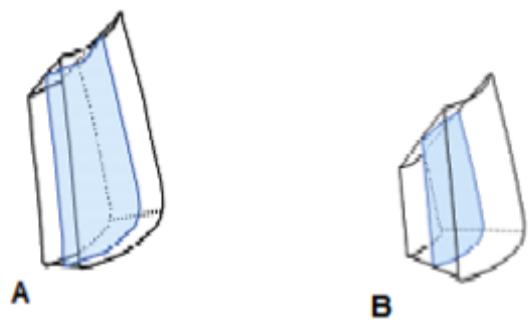

**Figure 1** A) demonstrates a myocardial segment at the end of diastole. B) demonstrates the same myocardial segment at the end of systole.

The data set that covers the entire left ventricle consists of short-axis (SA) images captured from the heart of 70 volunteers. The vector position of spatially designated muscle volume elements was mapped with a starting time at end diastole and time steps 70 ms apart. These positions were captured and registered relative to the positions of these elements at the end systole. The end systole time was determined through echocardiographic images and for this subject was 380 ms after QRS complex. The imaging parameters are as follows: repetition time =3.1 ms, mixing time =250 ms, flip angle = 90°, in-plane displacement encoding strength = 6.25 mm/$\pi$ , out-of-plane displacement encoding strength = 3.21 mm/$\pi$ , number of averages =3, number of phases = 3, in-plane resolution = 1.5 * 1.5$mm^2$ , slice thickness =5mm, and distance factor = 50%. Using the MATLAB view software, the three-dimensional force vector field was generated in a matrix format. Segmentation was then performed by masking all parts of the anatomy except for the myocardium. For this study, we have masked regions of the heart outside the left ventricle. Phase unwrapping was then performed on the segmented images by scanning the myocardium area while searching for sudden changes in the force magnitude. These phase wrappings were later unwrapped by adding or sub- tracting the force correction vector value, which corresponds to the $2\pi$ radiant changes in phase. This step was repeated separately for all three directions of force vector; MATLAB was used for the calculations. In (Figure 2), we can present arbitrary numbers SA slices of the heart at end systole along with arrows that show the force vector during the contraction that spans from end diastole to end systole. [33] By having data points across the left ventricle wall, depending on the wall thickness, we are able to calculate the transmural changes of the thickening and shortening index across the wall [12−17] and orientation of vectors are better illustrated in (Figure 2), It should be noted that the direction of these vectors represents the force directions that resulted from the contraction of many myofibers. Therefore, vector directions are not necessarily aligned in the muscle fiber directions at each myocardium point. The reduction of the left ventricle volume in systole, and therefore its pumping function, is mostly caused by wall thickening [18], which is the effect of tangential shortening of the myocardium. Therefore, these two dependent quantities can be used as the quantitative characteristics for the local contribution of the left ventricle myocardium to global heart function. The spatial distribution of regions that contribute the most to cardiac function acts as a functional macrostructure for the myocardium. The mere existence of such a distinct structure and the knowledge of its normal morphology will facilitate a more effective modeling of left ventricle function.

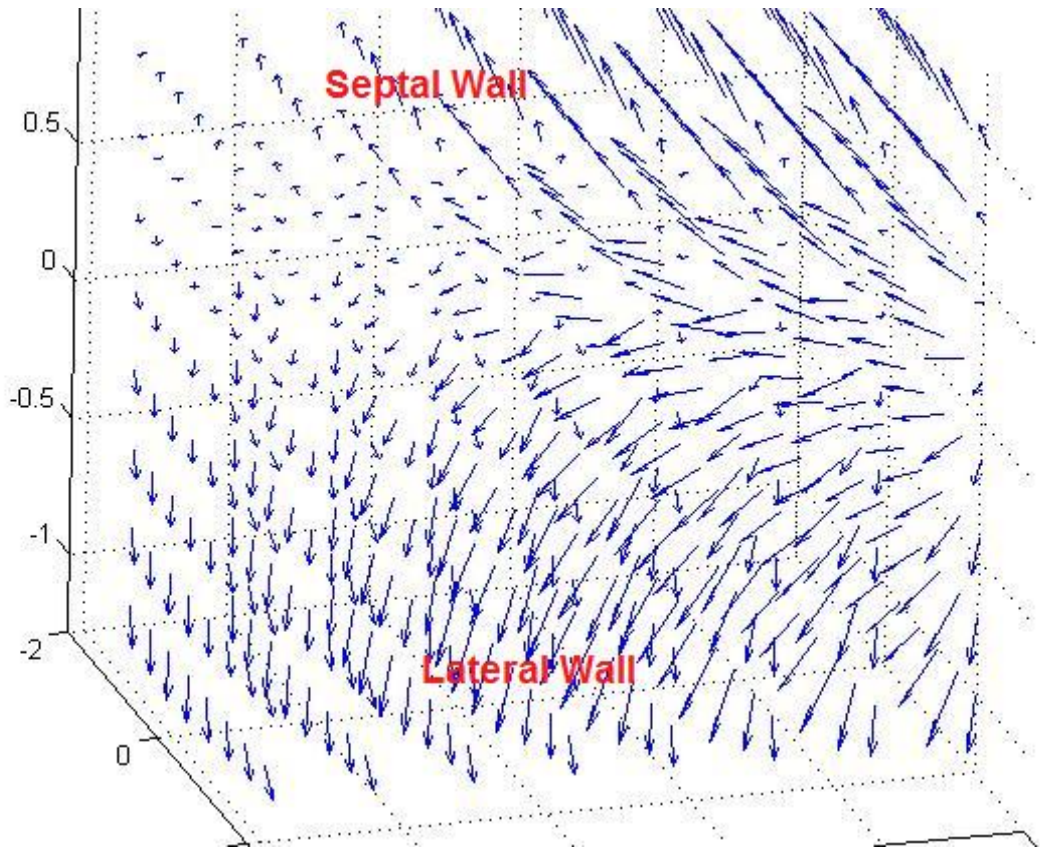

**Figure 2** The force vector field on a parasternal long-axis view in MATLAB software.

**Result:**

Assembled from data shown in (Figure 2), we can illustrate a segment of the left ventricle with tracking results from all areas through the heart wall from endocardium to epicardium (Figure 3).

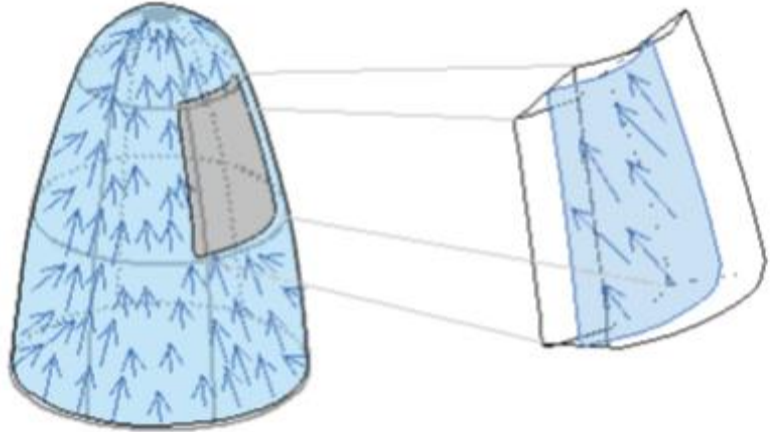

**Figure 3** Illustration of the averaged myocardial mesh model of the left ventricle. The arrows indicate the estimated force vector from one frame to the next.

We have suggested a scalar quantity to measure the tangential shortening in a manner that is not sensitive to the deformations in the tangent plane. [19,33] This quantity, known as the shortening index (SI), measures the change of the in-plane area based on the stretch tensor $U$, in the radial, longitudinal, and circumferential (RLC) coordinate system. With the consideration of the incompressibility of the myocardium, the local thickening of the wall (T) is also calculated by elements of U. Therefore, $T$ and SI can be measured at any point inside the heart muscle using stretch tensor in MATLAB software. These two dependent quantities have been used in this study as surrogates for regional contraction level. With the use of high resolution force field of the myocardium, maps of $T$ and SI with high spatial resolution were calculated based on the maximum deformation of the left ventricle, i.e., from the end diastole to the end systole. The macrostructure of the left ventricle was subsequently sought by calculation and comparison of the myofibers of these quantities for their relatively higher absolute values. The myofibers were calculated for SI values above a certain threshold. Starting from low values near zero, we gradually increased the magnitude of the threshold and removed myocardium regions that were left out of myofibers corresponding to the threshold. According to Moore et al. the maximum mean SI magnitude in the left ventricle is -0.25 0.05 in healthy hearts. [20] (Figure 4) shows myofiber transactions of left ventricle wall thickening and tangential shortening, corresponding to, assembled from data shown in (Figure 2) that myofibres move on helical bands which begin from the Septal, loop around the Apex and then go to the Anterior.

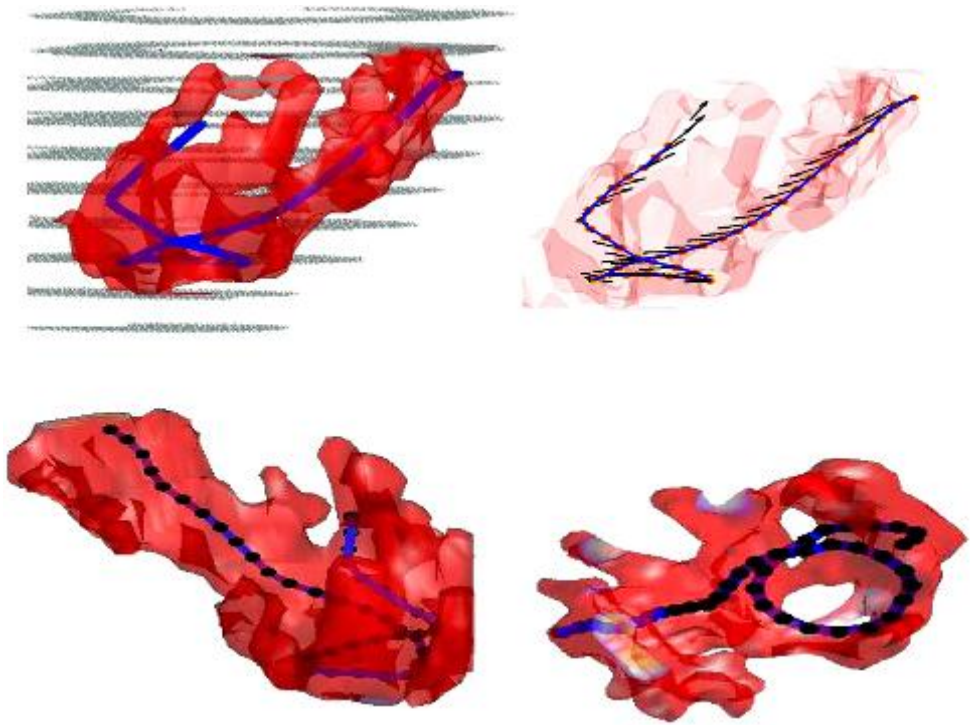

**Figure 4** The rout of a fiber in the left ventricle.

Finally, we are able to define the whole LV myocardial model mathematically, by MATLAB software in normal subjects (Figure 5).

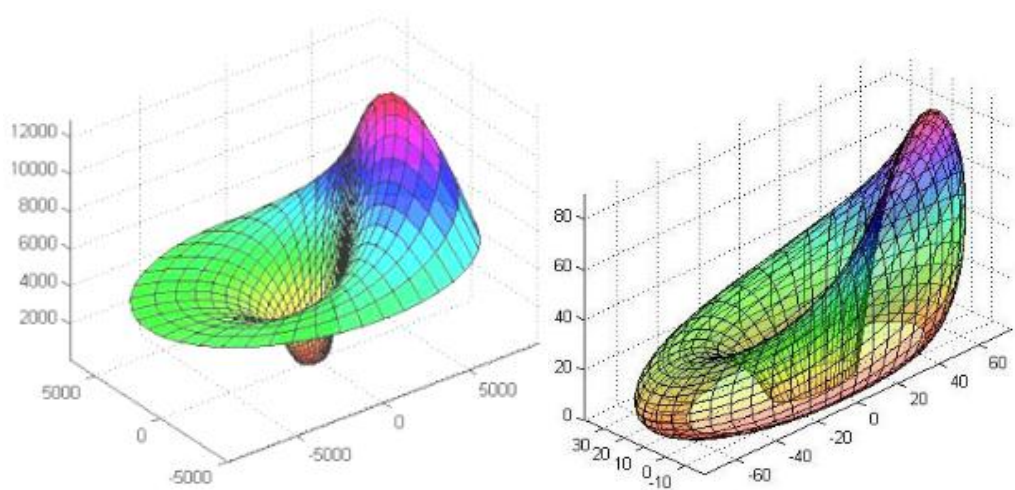

**Figure 6** Mathematical modeling of the left ventricle related to myocardial fiber paths in MATLAB software.

**Discussion:**

The development during this century, were concerned principally with the problems of existence of solutions and the integration of several broad categories of the boundary-value problems. A definitive treatment of fundamental problems in plane elasticity (primarily by the school of Russian mathematicians influenced by N. I. Muskhelishvili) was given and significant strides made in the theory of non-linear theories of elasticity. The theory of shell is still in the formative, patchy state characterized by conflicting approximations. A comprehensive treatment of the shell theory is given in the theory of thin Elastic Shells by A. L. Goldenveiser. [21,22]

By Heimdal A et al 2D Speckle tracking strain provides an objective way to quantify global and regional left ventricular (LV) systolic and diastolic function with improved accuracy and greater reproducibility. [23–25] And According to Luigi P. Badano, M.D. at the Department of Cardiac, Thoracic and Vascular Sciences, University of Padua, Italy, 4D strain has the potential to become the reference method to access myocardyal function and detect early, subclinical involvement in many heart diseases as well as to quantify regional myocardial function ischemic heart diseases. [26,27] These two investigations strongly use searching points by image processing machinery method but our studies demonstrate regional and global of the myocardial function by an introducing mathematical fibered modeled of it which under a software is very robust, reproducible and user friendly. Its clinical values remain to be established. Another Previous attempts to study the spatial distribution of the myofibers through blunt dissection [28] have suggested the existence of a helical myofiber arrangement interlaced with connective tissues within the ventricular mass. Recent imaging studies on the structure of the myocardium using diffusion tensor MRI (DTMRI) have produced evidence in support of the above mentioned conjecture. For example, works at University of Virginia and Johns Hopkins have clearly depicted a continuous bundle of muscle fibers in the helical form within the elaborate connective tissue matrix of the left ventricular mass. [29] It is important to note that in the DTMRI technique, the static structure of myofibers is mapped by following the diffusive direction of water molecules along these fibers in arrested hearts. Therefore, crucial dynamic information such as strain field must be obtained by other means. In this respect, it would not have been possible to confirm the functional role of such an intriguing structure in the cardiac function and dynamics. For this reason, these observations and other efforts to model cardiac dynamics based on the existence of a helical myofiber band have faced severe criticism due to the lack of evidence in support of the dynamical significance and functional role of a helical myofiber band within the chamber wall. [30] Our observation through DENSE imaging shows strong resemblance to the cardiac muscle fibers obtained by DTMRI as well as to the conjectures that originated from blunt dissection studies. [28,30] The main distinction is in the approaches used to identify and characterize the band. Unlike DTMRI and dissection techniques, we employed a dynamical characteristic approach to reach a similar morphological conclusion. When the function and the morphology of the most active region of the myocardium are linked, one can notice the ability of the heart to twist during the systole and untwist during diastole. [31,32] From a macromechanical point of view, such twisting dynamics cannot be produced without the presence of a helically oriented and dynamically preferential muscle morphology within the myocardium mass. It is natural to conclude that transmission of forces along such a helical pathway would generate global resultant motion of the chamber wall in the form of twisting and untwisting action, which has been observed and could not be fully explained by the prevailing localized, fiber structure models of the left ventricle. [31] As an application, when the pathological changes in the motion of the heart muscle fibers occurs it can be easily diagnosed related pathology. And meanwhile for the diagnosis of various pathologies requiring high experience in echocardiography is waved. The standard way to model the motion of blood inside the left ventricle would be to treat the left ventricle as an elastic membrane obeying Newton's laws of motion with forces calculated in part from the elasticity of the membrane and in part by evaluating the fluid stress tensor on the surface of the membrane. [33] Then the fluid equations would have to be supplemented by the constraint that the velocity of the fluid on either side of the membrane must agree with the instantaneously known velocity of the elastic membrane itself. There is a difficulty with this standard approach to the problem. Challenge is the practical one of evaluating the fluid stress tensor on either side of the boundary. This seems difficult (or at least messy) to do numerically, unless the computational grid is aligned with the boundary. On the other hand, in a moving boundary problem, it is both expensive and complicated to re-compute the grid at every time step in order to achieve alignment. This means that the sum of the elastic force and the fluid force on any part of the boundary has to be zero. Once we know this, it becomes unnecessary to evaluate the fluid stress tensor at the boundary at all! We can find the force of any part of the boundary on the fluid by evaluating the elastic force on that part of the boundary. (Note the use of Newton's third law: the force of boundary on fluid is minus the force of fluid on boundary (All we need is a method for transferring the elastic force from the boundary to the fluid. On a Cartesian grid, this may be done by spreading each element of the boundary force out over nearby grid points. The particular way that this is done in the boundary method involves a carefully constructed approximation to the Dirac delta function. This force-spreading operation defines a field of force on the Cartesian lattice that is used for the fluid computation. Then the fluid velocity is updated under the influence of that force field. The Navier-Stokes solver that updates the fluid velocity does not know about the any consideration of the heart left ventricle geometry; it just works with a force field that happens to be zero everywhere except in the immediate of the vorticity region. This approach can be used for instructional purposes and diagnosis of heart ailments. [34,35]

**Conclusion:**

In this paper, the relation of the form and function of the left ventricle (myocardium) has been studied through investigating the spatial distribution of its regional function. In this approach, we were able to define the whole LV myocardial model mathematically, by MATLAB software in normal subjects. This will enable physicians to diagnose and follow up many cardiac diseases when this software is interfaced within echocardiographic machines.

**Conflict of interest:**

There is no conflict of interest.